\documentstyle[aps,preprint,epsfig]{revtex}

\tightenlines

\begin{document}

\title{ Statistical Coalescence Model Analysis of\\
$J/\psi$ Production in Pb+Pb Collisions at 158~A$\cdot$GeV}

\author{
A.P. Kostyuk$^{a,b}$,
M.I. Gorenstein$^{b}$,
H. St\"ocker$^{a}$,
and
W. Greiner$^{a}$
}

\address
{$^a$ Institut f\"ur Theoretische Physik, Universit\"at  Frankfurt,
Germany}

\address{$^b$ Bogolyubov Institute for Theoretical Physics,
Kyiv, Ukraine}

\date{\today}
\maketitle

\begin{abstract}
Production of $J/\psi$ mesons in heavy ion collisions is considered within
the statistical coalescence model.
The model is in agreement with the experimental data of the NA50
Collaboration
for Pb+Pb collisions at 158~A$\cdot$GeV in a wide centrality range,
including the so called
``anomalous'' suppression domain.
The model description of the $J/\psi$ data requires, however, strong
enhancement of the open charm production
in central Pb+Pb  collisions. This model prediction may be checked
in the future SPS runs.
\end{abstract}

Production of charmonium states $J/\psi$ and $\psi^{\prime}$ in
nucleus-nucleus collisions has been studied at CERN SPS over the previous 15
years by the NA38 and NA50 Collaborations. This experimental program was
mainly motivated by the suggestion \cite{MS} to use the $J/\psi$ as a probe
of the state of matter created at the early stage of the collision. The
original picture \cite{MS} (see also \cite{Satz} for a modern review)
assumes that charmonia are created exclusively at the initial stage of the
reaction in primary nucleon-nucleon collisions. During the subsequent
evolution of the system, the number of hidden charm mesons is reduced
because of: (a) absorption of pre-resonance charmonium states by
nuclear nucleons (normal nuclear suppression), (b) interactions of
charmonia with secondary hadrons (comovers), (c) dissociation of
$c\bar{c}$ bound states in deconfined medium (anomalous suppression). It
was found \cite{NA38} that $J/\psi$ suppression with respect to Drell-Yan
muon pairs measured in proton-nucleus and nucleus-nucleus collisions with
light projectiles can be explained by the so called {\it ''normal''} (due
to sweeping nucleons) nuclear suppression alone. In contrast, the NA50
experiment with a heavy projectile and target (Pb+Pb) revealed
essentially stronger $J/\psi$ suppression for central collisions
\cite{anomalous,xsections,threshold,evidence}. This {\it anomalous}
$J/\psi$ suppression was attributed to formation of quark-gluon
plasma (QGP) \cite{evidence}, but a comover scenario cannot be excluded
\cite{comover}.

A completely different picture of charmonium production was developed
recently within several model approaches
\cite{GG,Br1,Go:00,Le:00,Ka:00,Ra:00}.
In contrast to the standard approach, hidden
charm mesons are supposed to be created at the hadronization stage of the
reaction due to coalescence of $c$ and $\bar{c}$ quarks created earlier.
In this case the $J/\psi$ yield is not restricted from above by the normal
nuclear suppression curve. Therefore, neither anomalous suppression nor
enhancement are excluded.

In the present letter we consider the statistical
coalescence model (SCM) \cite{Br1,Go:00} of charmonium
production. We assume that $c$ and $\bar{c}$ are created at the
initial stage of the reaction in primary hard parton collisions.
We neglect creation of $c\bar{c}$ pairs after the hard initial stage
as well as their possible annihilation. Then, the number of charmed
quark-antiquark pairs remain approximately unchanged during
subsequent stages. They are distributed over final hadron states
at the hadronization stage in accord with laws
of statistical mechanics.
The SCM provides an excellent quantitative description of the NA50
data on centrality dependence of $J/\psi$
production in Pb+Pb collisions at SPS, provided that the number
of nucleon participants is not too small ($N_p \agt 100$).
The peripheral collision data can be explained qualitatively.

If creation of heavy quarks is indeed a hard process only, the
average number $\langle c\bar{c} \rangle_{AB(b)}$ of produced
$c\bar{c}$ pairs must be proportional to the number of primary
nucleon-nucleon collisions.
Then the centrality dependence of $\langle c\bar{c} \rangle_{AB(b)}$ can be
calculated in Glauber's approach:
\begin{equation}
\langle c\bar{c} \rangle_{AB(b)} = \sigma^{NN}_{c\bar{c}} T_{AB}(b).
\end{equation}
Here $b$ is the impact parameter,
$T_{AB}(b)$ is the nuclear overlap function (see  Appendix) and
$\sigma^{NN}_{c\bar{c}}$ is the $c\bar{c}$ production cross section
for nucleon-nucleon collisions. As discussed in Ref.\cite{hf_enh},
deconfined medium can substantially modify charm production
in hard collisions at SPS.  Therefore,
$\sigma^{NN}_{c\bar{c}}$ in A+B collisions can be different from
the corresponding cross section measured in a nucleon-nucleon collision
experiment. The present analysis considers $\sigma^{NN}_{c\bar{c}}$ as
a free parameter. Its value is fixed by fitting the NA50 data.

Event-by-event fluctuations of the number of $c\bar{c}$ pairs follow
the binomial distribution, which can be safely approximated by the
Poisson distribution because the probability to produce a
$c\bar{c}$ pair in a nucleon-nucleon collision is small:
\begin{equation}
P_k(b) = \exp \left[ - \langle c\bar{c} \rangle_{AB(b)} \right]
\frac{[\langle c\bar{c} \rangle_{AB(b)}]^k}{k!},
\end{equation}
where $P_k(b)$ is the probability to produce $k$ $c\bar{c}$ pairs in
an A+B collision at impact parameter $b$.
Assuming exact $c\bar{c}$-number conservation during the
evolution of the system, the SCM result for the average number of produced
$J/\psi$ per A+B collision is given by \cite{Go:00}
\begin{equation}\label{Jpsi}
\langle J/\psi \rangle_{AB(b)} =
\langle c\bar{c} \rangle_{AB(b)}
\left[ 1 + \langle c\bar{c} \rangle_{AB(b)} \right]
\frac{N_{J/\psi}^{tot}}{(N_O/2)^2} +
o \left [ \frac{N_{J/\psi}^{tot}}{(N_O/2)^2}
\right]
.
\end{equation}
Here
\begin{equation}
N_O = \sum_{j=D,\bar{D},D^*,\bar{D}^*,\dots} N_j
\end{equation}
is the total open charm {\it thermal} multiplicity.
The sum runs over all known (anti)charmed particle species \cite{pdg}.
The total $J/\psi$ multiplicity includes the contribution of
excited charmonium states decaying into $J/\psi$:
\begin{equation}\label{dec}
N_{J/\psi}^{tot}= \sum_{j=J/\psi,\chi_1,\chi_2,\psi'}
R(j) N_j.
\end{equation}
Here $R(j)$  is the decay branching ratio of the  charmonium $j$
into $J/\psi$:
$R(J/\psi) \equiv 1$, $R(\chi_1)\approx 0.27$, $R(\chi_2)\approx 0.14$
and $R(\psi^{\prime})\approx 0.54$.
The multiplicities $N_j$ are found in the grand canonical ensemble
formulation of the equilibrium hadron gas model:
\begin{equation}\label{gce}
N_j~=~ V n_j(T,\mu_B) ~=~
\frac{d_j
V~e^{\mu_j/T}}{2\pi^2}~T~m_j^2~K_2\left(\frac{m_j}{T}\right)~.
\end{equation}
Here $V$ and $T$ are the volume\footnote{We use ideal HG formulas and
neglect excluded volume corrections.} and temperature of the HG system,
$m_j$ and $d_j$ denote, respectively, the masses and degeneracy
factors of particles. $K_2$ is the modified Bessel function.
The chemical potential $\mu_j$ of the particle species $j$ in Eq.(\ref{gce})
is defined as
\begin{equation}\label{mui}
\mu_j~=~b_j\mu_B~+~s_j\mu_S~+~c_j\mu_C~.
\end{equation}
Here $b_j$,$s_j$ and $c_j$ represent the baryon number, strangeness and
charm of the particle $j$, respectively. The baryonic chemical potential $\mu_B$
regulates the baryonic density. The
strange $\mu_S$ and charm $\mu_C$ chemical potentials are found
by requiring zero value for the total strangeness and charm
in the system . In our consideration we neglect small effects
of a non-zero electrical chemical potential.

We assume that the chemical freeze-out occurs close to (or even coincide
with) the hadronization (where charmonia supposedly formed).
Therefore for the thermodynamic parameters $T$ and
$\mu_B$ we use the chemical freeze-out values found \cite{Br2}
by fitting the  HG model to the hadron yield data in Pb+Pb collisions
at SPS:
\begin{equation}\label{set_Br}
T~=~168~{\rm MeV},~~ \mu_B~=~266~{\rm MeV}~.
\end{equation}
Uncertainties in the freeze-out parameters exist due to time evolution
of the system through the phase transition \cite{Spieles} and
because of possible change of
effective hadron masses in hot and dense hadron medium
\cite{Zschiesche}. To check robustness of
the predictions, an independent parameter set \cite{Becattini} is also
used. It has been obtained by assuming strangeness and antistrangeness
suppression by factor $\gamma_s$:
\begin{equation}\label{set_Bec}
T~=~158~{\rm MeV},~~ \mu_B~=~238~{\rm MeV}~~ \gamma_s~=~0.79~.
\end{equation}

The system is assumed to freeze-out chemically at some common volume. This
is fixed by the condition of baryon number conservation:
\begin{equation}\label{v_fix}
N_p(b)=V n_B(T,\mu_B) = V \sum_{j=N,\bar{N},\Delta,\bar{\Delta},\dots}
b_j n_j(T,\mu_B),
\end{equation}
Here $N_p(b)$ is the number of participating nucleons at impact parameter
$b$, $n_B(T,\mu_B)$ is the net baryon density at the chemical freeze-out.
The sum in Eq.(\ref{v_fix}) runs over all (anti)baryon species.
The linear relation (\ref{v_fix}) between $N_p$ and $V$ as well as
constant values of $T$ and $\mu_B$ parameters are assumed for all
collisions with different values of impact parameter $b$.

Eq. (\ref{Jpsi}) gives the {\it total} number of produced $J/\psi$-s. They
decay into $\mu^+\mu^-$ with the probability $B^{J/\psi}_{\mu\mu}= (5.88
\pm 0.10) \%$ \cite{pdg}. Only the fraction $\eta$ of $\mu^+\mu^-$ pairs that
satisfies the kinematical conditions
\begin{eqnarray}
&& 0 < y < 1, \label{rapidity} \\
&& -1/2 < \cos \theta < 1/2 \label{angle}
\end{eqnarray}
can be registered by the NA50 spectrometer. Here $y$ stands for the rapidity
of a $\mu^+\mu^-$ pair in the center-of-mass frame of colliding
nuclei. $\theta$ is the polar angle of the muon momentum
in the rest frame of the pair.
An estimate of $\eta$ is impossible without detailed information about
the hydrodynamic expansion of the system and the conditions at the thermal
freeze-out. We shall therefore treat $\eta$ as one more free parameter.

In the NA50 experiment the Drell-Yan muon pair multiplicity (either measured
or calculated from the minimum bias data) is used as a reference for
the $J/\psi$ suppression pattern. Similarly to $c\bar{c}$ pairs,
the number Drell-Yan pairs is proportional to the number of primary
nucleon-nucleon collisions:
\begin{equation}
\langle DY' \rangle_{AB(b)} = \sigma^{NN}_{DY'} T_{AB}(b),
\end{equation}
where $\sigma^{NN}_{DY'}$ is the nucleon-nucleon
production cross section of $\mu^+ \mu^-$
Drell-Yan pairs. The prime means that the pairs should satisfy
the kinematical conditions of the NA50 spectrometer
(\ref{rapidity}) and (\ref{angle}).
As the Drell-Yan cross section is isospin dependent, an average value
is used:
\begin{equation}
\sigma^{NN}_{DY'}=\frac{\sigma^{AB}_{DY'}}{AB}.
\end{equation}
For the case of Pb+Pb collisions,
$A=B=208$ and
$\sigma^{PbPb}_{DY'} = 1.49 \pm 0.13 $ $\mu$b \cite{xsections}.

Hence, the quantity to be studied is the ratio
\begin{equation}\label{Rb}
R(b) = \frac{\eta B^{J/\psi}_{\mu\mu} \langle J/\psi \rangle_{AB(b)}}
{\langle DY' \rangle_{AB(b)}}
=
\eta B^{J/\psi}_{\mu\mu}
\frac{\sigma^{NN}_{c\bar{c}}}{\sigma^{NN}_{DY'}}
\left( 1 +  \sigma^{NN}_{c\bar{c}} T_{AB}(b) \right)
\frac{N_{J/\psi}^{tot}}{(N_O/2)^2} \ .
\end{equation}
It is convenient to rewrite the last expression in a simpler form
\begin{equation}\label{Rbc}
R(b)=C \frac{1 +  \sigma^{NN}_{c\bar{c}} T_{AB}(b) }{N_p(b)}
\end{equation}
and treat $C$ and $\sigma^{NN}_{c\bar{c}}$ as free parameters.
In this form our fitting procedure does not depend on chemical
freeze-out conditions.
The new free parameter C is connected to $\eta$ by the expression
\begin{equation}
C = \eta B^{J/\psi}_{\mu\mu} \
\frac{\sigma^{NN}_{c\bar{c}}}{\sigma^{NN}_{DY'}} \
\frac{n_{J/\psi}^{tot}(T,\mu_B) n_B(T,\mu_B)}{(n_O(T,\mu_B)/2)^2}.
\end{equation}
Here we have introduced the total open (anti)charm density:
$n_O=N_O/V$ and total $J/\psi$ ``density'':
$n_{J/\psi}^{tot} = N_{J/\psi}^{tot}/V$. The relation between $C$
and $\eta$ does depend on freeze-out conditions, but our
calculations with the parameter sets (\ref{set_Br}) and (\ref{set_Bec})
have shown that this dependence is not essential.

In the NA50 experiment, the neutral transverse energy of produced particles
$E_T$ was used to measure centrality of the collisions. This variable,
however, provides a reliable measure of the centrality only if it does
not exceed a certain maximum value: $E_T \alt 100$ GeV
(see also Ref.\cite{Kaidalov,Dinh}). To show this
we have calculated the dependence of the average number of participants
on the transverse energy $\overline{N_p}(E_T)$.

The conditional probability to measure some value of $E_T$ at fixed
impact parameter $b$ is given by a gaussian distribution:
\begin{equation}
P(E_T|b) = \frac{1}{\sqrt{2 \pi q^2 a N_p(b)}} \exp \left( - \frac{[E_T -
q N_p(b)]^2}{2 q^2 a N_p(b)} \right).
\end{equation}
Analyzing the experimental situation, we are interested in a quite opposite
question: how events with fixed $E_T$ are distributed with respect
to the centrality. The answer is
\begin{equation}
P(b|E_T) = \frac{b P(E_T|b) P_{int}(b)}
{\int_0^{+\infty} db b P(E_T|b) P_{int}(b)},
\end{equation}
where $P_{int}(b)$ stands for the probability (see Appendix) that two nuclei
at fixed impact parameter $b$
interact (at least one pair of nucleons collides).
The average number of participating nucleons at fixed $E_T$ is then
given by the expression:
\begin{equation}
\overline{N_p}(E_T) =
\int_0^{+\infty} db N_p(b) P(b|E_T)
=
\frac{\int_0^{+\infty} db b N_p(b) P(E_T|b) P_{int}(b)}
{\int_0^{+\infty} d b b P(E_T|b) P_{int}(b)}.
\end{equation}
The parameter values $q=0.274$ GeV and $a=1.27$ \cite{Chaurand} are
fixed from the minimum bias transverse energy distribution.

\begin{figure}[p]
\begin{center}
\vfill
\epsfig{file=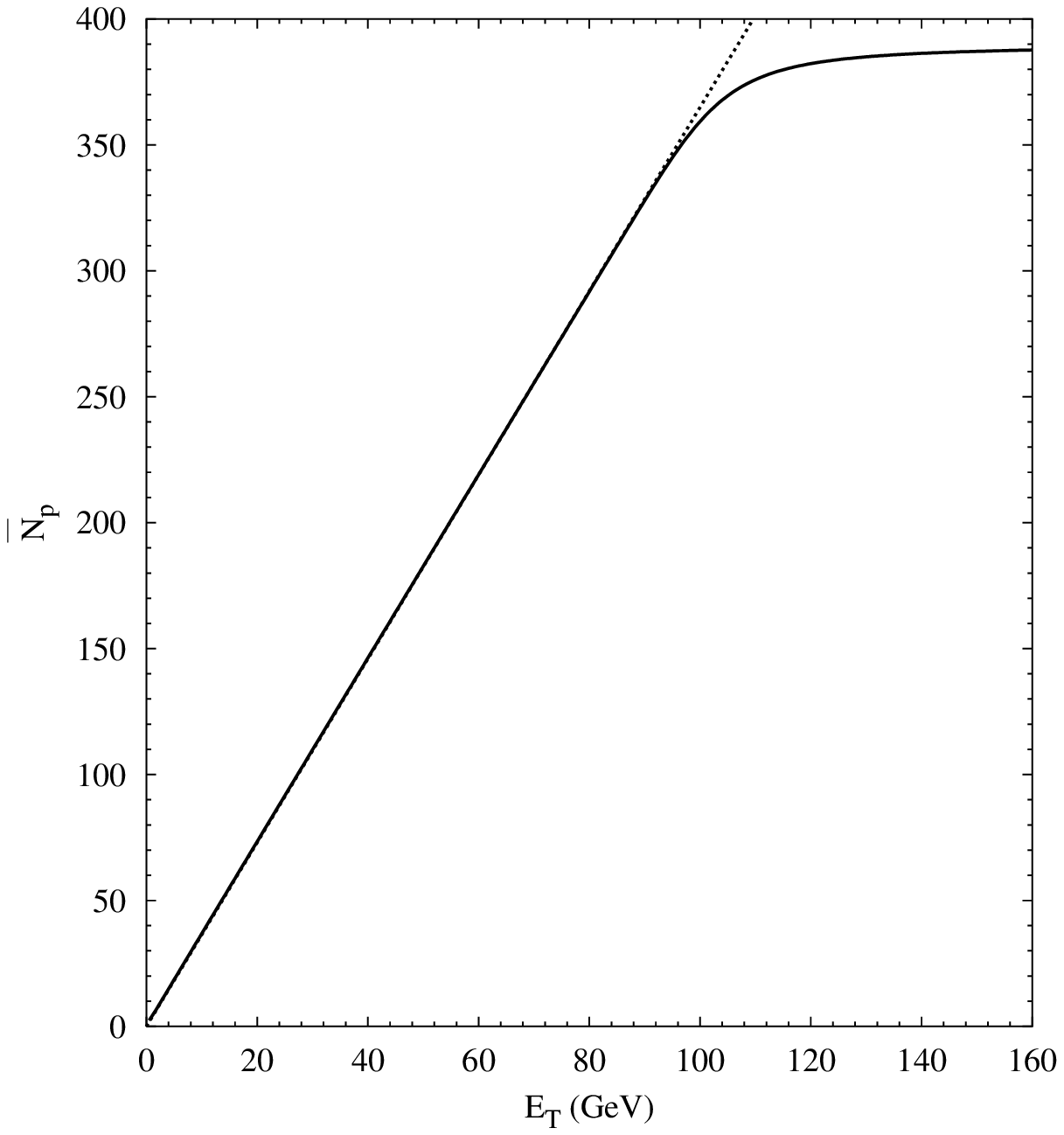,height=17cm}
\mbox{}\\
\vfill
\caption{The dependence of the average number of nucleon participants on
the transverse energy. The dotted straight line corresponds to
Eq.(\ref{linear}).
\label{NpET}
}
\end{center}
\end{figure}

The result is shown in Fig.\ref{NpET}. As is seen, the transverse energy
is simply related to the number of participants $E_T = q \overline{N_p}$
in the domain $E_T \alt 100$ GeV. Outside of this domain $\overline{N_p}$
does not change essentially as $E_T$ grows. Therefore the data at $E_T >
100$ GeV do not represent centrality dependence of the $J/\psi$
suppression pattern but rather its dependence on fluctuations of the
stopping energy at {\it fixed} number of participants. In principle,
influence of such fluctuations on $J/\psi$ multiplicity can be studied in
the framework of our model, but information concerning the
corresponding fluctuations of the chemical freeze-out parameters $T$ and
$\mu_B$ would be needed. Experimental data that would allow to
extract this information (hadron yields at extremely large transverse
energy) are not available at present. Therefore, we restrict our analysis
to centrality dependence of $J/\psi$ production and do not use the
data corresponding to large transverse energies $E_T > 100$ GeV.

On the other hand, the SCM is not expected to describe small systems.
This can be seen from $\psi'$ data \cite{Br1}.
In the framework of SCM the multiplicity of
$\psi'$ is given by the formula (\ref{Jpsi}) with the replacement
$N_{J/\psi}^{tot} \rightarrow N_{\psi'}$. Therefore, the $\psi'$ to
$J/\psi$ ratio as a function of centrality should be constant and equal
to its thermal equilibrium value. The experimental
data \cite{Jaipur} (see also a compilation in Ref. \cite{Br1})
are consistent with this picture only at rather large
($N_p \agt 100$) numbers of participants  \cite{Shuryak}.

\begin{figure}[p]
\begin{center}
\vfill
\epsfig{file=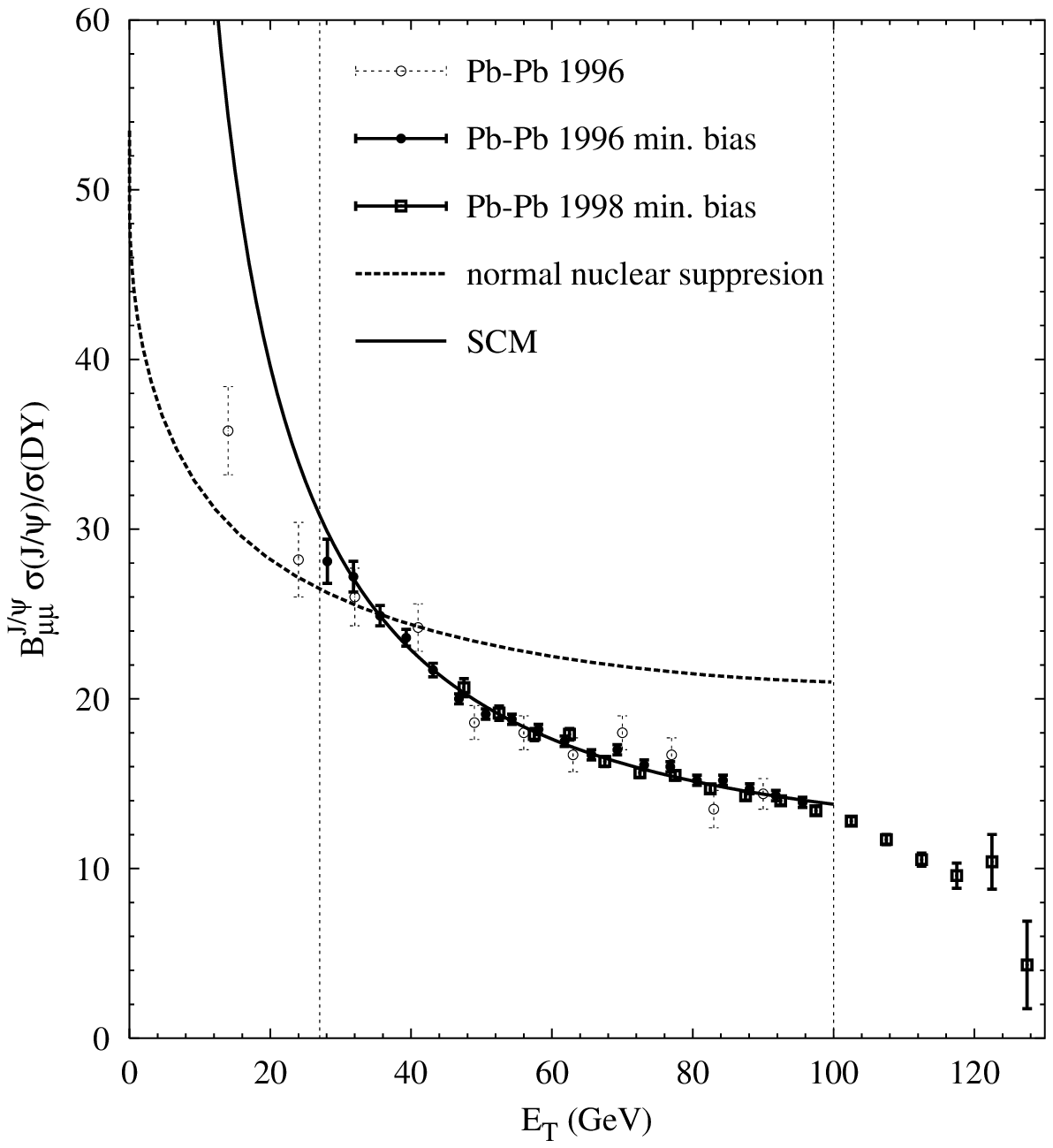,height=17cm}
\mbox{}\\
\vfill \caption{The dependence of $J/\psi$ over Drell-Yan multiplicity
ratio on the transverse energy. The normal nuclear suppression curve is
obtained at $\sigma_{abs} = 6.4$~mb, where $\sigma_{abs}$ is the absorption
cross section of preresonant charmonia by nuclear nucleons.
Two vertical lines show the applicability domain of the model under
consideration, see the text for details.
\label{f1} }
\end{center}
\end{figure}

Hence, the applicability domain of the
model is limited to
\begin{equation}\label{apldom}
27 < E_T < 100 \mbox{ GeV.}
\end{equation}
Note that the most
precise and abundant NA50 data (see Fig. \ref{f1}) correspond to this
kinematical region.

At $E_T \alt 100$ GeV the formula (\ref{Rbc}) and the
equation
\begin{equation}\label{linear}
E_T = q N_p(b)
\end{equation}
give a parametric dependence of the ratio $R$ on the transverse
energy. This dependence for the parameter set
\begin{eqnarray}
C &=& (2.59 \pm 0.25) \cdot 10^3  \nonumber \\
\sigma^{NN}_{c\bar{c}} &=& (34 \pm 10) \mbox{ $\mu$b}
\label{parset}
\end{eqnarray}
is plotted in Fig. \ref{f1}. The free parameters were fixed by fitting
three sets of NA50 data \cite{threshold,evidence} within the applicability
domain (\ref{apldom}) of the model  by the least
square method. The model demonstrates excellent agreement with the
fitted data ($\chi^2/\mbox{dof} = 1.2$).

Extrapolation of the fit to peripheral collisions reveals
sharp increase
of the ratio (\ref{Rb}) with decreasing $N_p$.
Such behavior in the SCM can be understood as the following.
The smaller is the volume of the
system the larger is the
probability that $c$ and $\bar{c}$ meet each other at hadronization
stage and form a hidden charm meson.
As is seen from Fig.~\ref{f1}, this is not
supported by the data: the SCM  curve
lies above the experimental points in the low $E_T$ region.
On the other hand, the normal nuclear suppression model also fails
to explain the leftmost point from the 1996 standard analysis set
and two leftmost points from the 1996 minimum bias set. Those theoretical
calculations underestimate the experimental
values.  It is natural to assume that an intermediate situation
takes place\footnote{Similar combination of standard and SCM production
mechanisms has been considered in Ref.\cite{Rapp} for central Pb+Pb
collisions. It was not, however, checked whether this approach is able
to describe the centrality dependence of the $J/\psi$ suppression
pattern in (semi)central collision region.}. Some fraction of 
peripheral Pb+Pb collisions result in
formation of deconfined medium. In these collisions charmonia are
formed at the hadronization stage, and their multiplicities are given
by SCM. The rest collisions (we shall call them 'normal collisions')
do not lead to color deconfinement, therefore charmonia are formed
exclusively at the initial stage and then suffer
normal nuclear suppression.
The experiment measures the average value, which lies between the
two curves.

The fraction of 'normal' events decreases with growing centrality.
Their influence on $J/\psi$ production becomes negligible
at $N_p \agt 100$. To check this we repeated the above fitting
procedure using only the experimental data corresponding to
$N_p > 200$.
The quality of the fit is only slightly better: $\chi^2/\mbox{dof} = 1.1$,
the parameter values
$C= (2.73 \pm 0.40) \cdot 10^3$  and
$\sigma^{NN}_{c\bar{c}} = (31 \pm 12)$ $\mu$b
are consistent with the analysis of the full
data set (\ref{parset}).

Our picture is also supported by $\psi'$ data. The normal nuclear
suppression influence nascent charmonia before the formation of meson
states. Therefore its effect on $\psi'$ is the same as on $J/\psi$. The
multiplicity ratio of $\psi'$ to $J/\psi$ in 'normal' nuclear-nuclear
collisions should be the same as in nucleon-nucleon collisions and should
not depend on the centrality. In the framework of SCM, the $\psi'$ to
$J/\psi$ ratio, as was explained above, should be equal to its thermal
equilibrium value, which is a few times smaller than the corresponding
value for 'normal' collisions. As the
fraction of 'normal' events decreases, the measured ratio should decrease
and then become constant and equal to its thermal value. The experimental
data  \cite{Jaipur} indeed demonstrate such behavior \cite{Shuryak,Br1}.

The present analysis predicts strong enhancement of the total
number of charm. From a pQCD fit of available data on charm
production in p+N and p+A collisions, one could expect
$\sigma^{NN}_{c\bar{c}} \approx 5.5$~$\mu$b at $\sqrt{s}=17.3$ GeV.
Our result (\ref{parset}) is larger by a factor of $4.5 \div 8.0$,
which is around the upper bound of the charm enhancement
estimated in Ref.\cite{hf_enh}.

Formation of deconfinement medium can change not only the total number
of charmonia and open charm particles but also their rapidity
distributions. For direct charmonium production in hard parton collisions,
dimuon pairs satisfying the kinematical conditions (\ref{rapidity})
and (\ref{angle}) account for a fraction of about $\eta_{hard} \approx
0.24$ in the total number of pairs originating from $J/\psi$ decays.
(The value was found using Schuler's parameterization \cite{Schuler}.)
Our result (\ref{parset}) corresponds to $\eta \approx 0.14$, which is by
a factor of about $0.6$ smaller. This difference can be attributed to
broadening of the $J/\psi$ rapidity distribution. It is natural to expect
similar modification of the open charm rapidity distribution. Because of
this modification the open charm enhancement within a limited rapidity
window can, in general, differ from the one for the total phase space.
Assuming that the broadening  for the open charm is approximately the same
as that for $J/\psi$, one obtains open charm enhancement by a factor of
about $2.5 \div 4.5$ within the rapidity window (\ref{rapidity}), which is
consistent with the indirect experimental result
\cite{NA50open}\footnote{Our previous study \cite{Go:00} were based on the
$J/\psi$ multiplicity data \cite{Marek}, which were extracted from the
NA50 data \cite{threshold,evidence} assuming narrow rapidity
distribution of $J/\psi$. In this case, the charm enhancement in the total
phase space is by a factor of about $2$--$3.5$ and does not differ from the
enhancement in the limited rapidity domain, but the number of $c\bar{c}$
pairs should grow faster than the number of nucleon-nucleon collisions.
Either data on rapidity distribution of $J/\psi$ or precise data
on centrality dependence of the open charm multiplicity would help us
to decide, which of these two versions is preferable. }.

In conclusion, we have shown that the NA50 data on centrality
dependence of the $J/\psi$ and $\psi'$ production in Pb+Pb collisions
\cite{threshold,evidence,Jaipur} are consistent with the following scenario:

The deconfined medium, which is formed in a Pb+Pb collision, prevents
formation of charmonia at the initial stage of the reaction. Instead,
hidden charm mesons are created at the hadronization stage due to
coalescence of created earlier $c$ and $\bar{c}$ quarks. Within this
scenario, the color deconfinement does not necessary lead to suppression
of $J/\psi$. Both suppression and enhancement are possible
\cite{Go:02}. If the number
of nucleon participants is not too small ($N_p \agt 100$), the number of
produced $J/\psi$ is smaller than in the case of normal nuclear
suppression, therefore {\it anomalous suppression} is observed. As
color deconfinement is present in most collision events for $N_p
\agt 100$, our model reveals excellent agreement with the experimental
data in this centrality domain.
The statistical coalescence model does not describe the NA50 data for
the peripheral Pb+Pb collisions.
It seems that the fraction of events
producing the deconfinement medium
is not dominating there and most of peripheral collisions follow the
normal nuclear suppression
scenario. Still, the presence a fraction of abnormal events could reveal
itself in the deviation of the $J/\psi$ data up from the normal nuclear
suppression curve.

Our model analysis predicts rather strong enhancement of the open charm.
This effect can also be related to the color deconfinement \cite{hf_enh}.
The enhancement within the rapidity window $0 < y < 1$ is consistent
with the indirect NA50 data \cite{NA50open}. A direct measurement of the
open charm would be very important for checking the above scenario.

\acknowledgments
The authors
are thankful to F. Becattini, K.A.~Bugaev, P.~Bordalo,
M. Ga\'zdzicki, L.~Gerland, L.~McLerran and K.~Redlich  for comments and
discussions.
We acknowledge the financial support of
the Alexander von Humboldt Foundation, Germany.
The research described in this publication was made possible in part
by Award \# UP1-2119 of the U.S. Civilian Research and Development
Foundation for the Independent States of the Former Soviet Union
(CRDF) and INTAS grant 00-00366.

\appendix

\section*{Nuclear geometry}

The spherically symmetrical distribution of nucleons in
the Pb-208 nucleus can be parameterized by a two-parameter
Fermi function \cite{tables} (this parameterization is also known
as the Woods-Saxon distribution):
\begin{equation}\label{2pF}
\rho(r)=\rho_0 \left[1+exp\left(
\frac{r-c}{a}
\right)\right]^{-1}
\end{equation}
with $c \approx 6.624$ fm, $a \approx 0.549$ fm and $\rho_0$ is
fixed by the normalization condition:
\begin{equation}\label{norm_rho}
4 \pi \int_0^\infty dr r^2 \rho(r) = 1 .
\end{equation}

The nuclear thickness distribution $T_A(b)$ is given by the
formula
\begin{equation}\label{T_A}
T_A(b) = \int_{-\infty}^{\infty} d  z \rho \left( \sqrt{b^2 + z^2} \, \right)~,
\end{equation}
and the nuclear overlap function is defined as
\begin{equation}\label{T_AB}
T_{AB}(b)=\int_{-\infty}^{\infty} d  x
\int_{-\infty}^{\infty} d  y \
T_A \left( \sqrt{x^2+y^2} \, \right) T_B \left( \sqrt{x^2+(y-b)^2} \, \right)~.
\end{equation}
 From Eq.(\ref{norm_rho}), one can deduce that
the above functions satisfy the following normalization conditions:
\begin{equation}\label{norm_T_AB}
2 \pi \int_0^\infty d b \, b \, T_A(b) = 1~,~~~~
2 \pi \int_0^\infty d b \, b \, T_{AB}(b) = 1 .
\end{equation}

In Glauber's approach the average number of participants
(`wounded nucleons') in A+B collisions at impact parameter $b$
is given by \cite{Bialas}
\begin{eqnarray}\label{Npart0}
\tilde{N}_p(b) &=& A~\int_{-\infty}^{+\infty}  d x
\int_{-\infty}^{+\infty}  d y
~T_A \left( \sqrt{x^2+y^2} \right)
\left\{ 1 - \left[ 1-
\sigma_{N N}^{inel} T_B \left( \sqrt{x^2+(y-b)^2} \right)
\right]^B \right\} \\
& &  +~
B~\int_{-\infty}^{+\infty}  d x
\int_{-\infty}^{+\infty}  d y
~T_B \left( \sqrt{x^2+(y-b)^2} \right)
\left\{ 1 - \left[ 1-
\sigma_{N N}^{inel} T_A\left( \sqrt{x^2+y^2} \right)
\right]^A \right\}~.  \nonumber
\end{eqnarray}
Here $\sigma_{N N}^{inel}$ is the nucleon-nucleon total inelastic
cross section.

At large impact parameter, the nuclei may do not interact at all.
Therefore $\tilde{N}_p(b) \rightarrow 0$ at $b \rightarrow \infty$.
If one interested in the average number of participants, provided
that an interaction between two nuclei has taken place, the relevant
quantity is
\begin{eqnarray}\label{Npart}
N_p(b) &=& \tilde{N}_p(b)/P_{int}(b),
\end{eqnarray}
where
\begin{equation}\label{Pint}
P_{int}(b) = 1 - \left[ 1-\sigma^{NN}_{inel} T_{AB}(b) \right]^{AB}.
\end{equation}
is the probability for nuclei A and B to interact at impact
parameter $b$. Although $N_p(b)$ differ from $\tilde{N}_p(b)$
at large $b$: $\tilde{N}_p(b) \rightarrow 2$ at
$b \rightarrow \infty$, they are almost identical for more central
collisions.

The average number of nucleon-nucleon collisions can be calculated
from
\begin{equation}
\tilde{N}_{coll}(b) = AB \sigma^{NN}_{inel} T_{AB}(b).
\end{equation}
Provided that an interaction between two nuclei has taken place,
the above formula should be modified as
\begin{equation}
N_{coll}(b) = \tilde{N}_{coll}(b)/P_{int}(b). \label{Ncoll}
\end{equation}

\widetext

\widetext


\begin{thebibliography}{99}

\bibitem{MS}
T.~Matsui and H.~Satz,
T.~Matsui and H.~Satz,
Phys.\ Lett.\ B {\bf 178} (1986) 416.

\bibitem{Satz}
H.~Satz,
Rept.\ Prog.\ Phys.\  {\bf 63} (2000) 1511
[hep-ph/0007069].

\bibitem{NA38}
M.~C.~Abreu {\it et al.},
Phys.\ Lett.\ B {\bf 466} (1999) 408.

\bibitem{anomalous}
M.~C.~Abreu {\it et al.}  [NA50 Collaboration],
Phys.\ Lett.\ B {\bf 410} (1997) 337.

\bibitem{xsections}
M.~C.~Abreu {\it et al.}  [NA50 Collaboration],
Phys.\ Lett.\ B {\bf 410} (1997) 327.

\bibitem{threshold}
M.~C.~Abreu {\it et al.}  [NA50 Collaboration],
Phys.\ Lett.\ B {\bf 450} (1999) 456.

\bibitem{evidence}
M.~C.~Abreu {\it et al.}  [NA50 Collaboration],
Phys.\ Lett.\ B {\bf 477} (2000) 28.

\bibitem{comover}
C.~Spieles, R.~Vogt, L.~Gerland, S.~A.~Bass, M.~Bleicher,
H.~Stocker and W.~Greiner,
Phys.\ Rev.\ C {\bf 60} (1999) 054901
[hep-ph/9902337];\\
J.~Geiss, C.~Greiner, E.~L.~Bratkovskaya, W.~Cassing and U.~Mosel,
Phys.\ Lett.\ B {\bf 447} (1999) 31
[nucl-th/9803008];\\
N.~Armesto, A.~Capella and E.~G.~Ferreiro,
Phys.\ Rev.\ C {\bf 59} (1999) 395
[hep-ph/9807258];\\
D.~E.~Kahana and S.~H.~Kahana,
Prog.\ Part.\ Nucl.\ Phys.\ {\bf 42} (1999) 269.

\bibitem{GG}
M.~Ga\'zdzicki and M.~I.~Gorenstein,
Phys.\ Rev.\ Lett.\  {\bf 83} (1999) 4009
[hep-ph/9905515].

\bibitem{Br1}
P.~Braun-Munzinger and J.~Stachel,
Phys.\ Lett.\ B {\bf 490}, (2000) 196 [nucl-th/0007059].

\bibitem{Go:00}
M.~I.~Gorenstein, A.~P.~Kostyuk, H.~Stocker and W.~Greiner,
Phys.\ Lett.\ B {\bf 509} (2001) 277
[hep-ph/0010148];\\
J.\ Phys.\ G {\bf 27} (2001) L47
[hep-ph/0012015].

\bibitem{Le:00}
P.~Csizmadia and P.~L\'evai,
hep-ph/0008195;\\
P.~L\'evai, T.~S.~Bir\'o, P.~Csizmadia, T.~Cs\"org\H o and J.~Zim\'anyi,
J.\ Phys.\ G {\bf G27} (2001) 703
[nucl-th/0011023].

\bibitem{Ka:00}
S.~Kabana,
J.\ Phys.\ G {\bf G27} (2001) 497 [hep-ph/0010228].

\bibitem{Ra:00}
R.~L.~Thews, M.~Schroedter and J.~Rafelski,
Phys.\ Rev.\ C {\bf 63} (2001) 054905
[hep-ph/0007323].

\bibitem{hf_enh}
A.~P.~Kostyuk, M.~I.~Gorenstein and W.~Greiner,
Phys.\ Lett.\ B {\bf 519} (2001) 207
[hep-ph/0103057].


\bibitem{pdg}
D.~E.~Groom {\it et al.}  [Particle Data Group Collaboration],
Eur.\ Phys.\ J.\ C {\bf 15} (2000) 1.

\bibitem{Br2}
P.~Braun-Munzinger, I.~Heppe and J.~Stachel,
Phys.\ Lett.\ B {\bf 465} (1999) 15
[nucl-th/9903010].

\bibitem{Spieles}
C.~Spieles, H.~Stocker and C.~Greiner,
Phys.\ Rev.\ C {\bf 57} (1998) 908
[hep-ph/9708280].

\bibitem{Zschiesche}
D.~Zschiesche, P.~Papazoglou, C.~W.~Beckmann, S.~Schramm,
J.~Schaffner-Bielich, H.~Stocker and W.~Greiner,
Nucl.\ Phys.\ A {\bf 663} (2000) 737
[nucl-th/9908072].

\bibitem{Becattini}
F.~Becattini, J.~Cleymans, A.~Keranen, E.~Suhonen and K.~Redlich,
hep-ph/0002267.

\bibitem{Kaidalov}
A.~Capella, E.~G.~Ferreiro and A.~B.~Kaidalov,
Phys.\ Rev.\ Lett.\  {\bf 85} (2000) 2080
[hep-ph/0002300].

\bibitem{Dinh}
J.~Blaizot, M.~Dinh and J.~Ollitrault,
Phys.\ Rev.\ Lett.\  {\bf 85} (2000) 4012
[nucl-th/0007020].

\bibitem{Chaurand}
B.~Chaurand, quoted in Ref. \cite{Dinh}.


\bibitem{Jaipur}
M.~Gonin {\it et al.},
PRINT-97-208
{\it Presented at 3rd International Conference on Physics and Astrophysics
of Quark Gluon Plasma (ICPAQGP 97), Jaipur, India, 17-21 Mar 1997}.

\bibitem{Shuryak}
H.~Sorge, E.~Shuryak and I.~Zahed,
Phys.\ Rev.\ Lett.\  {\bf 79} (1997) 2775
[hep-ph/9705329].

\bibitem{Rapp}
L. Grandchamp and R. Rapp, hep-ph/0103124.


\bibitem{Schuler}
G.~A.~Schuler,
CERN-TH-7170-94, hep-ph/9403387.

\bibitem{NA50open}
M.~C.~Abreu {\it et al.}  [NA38 and NA50 Collaborations],
Eur.\ Phys.\ J.\ C {\bf 14} (2000) 443.

\bibitem{Marek}
M.~Ga\'zdzicki,
Phys.\ Rev.\ C {\bf 60} (1999) 054903
[hep-ph/9809412].

\bibitem{Go:02}
M.~I.~Gorenstein, A.~P.~Kostyuk, L.~D.~McLerran, H.~Stoecker and W.~Greiner,
J.\ Phys.\ G, in print [arXiv:hep-ph/0012292];\\
M.~I.~Gorenstein, A.~P.~Kostyuk, H.~Stocker and W.~Greiner,
Phys.\ Lett.\ B {\bf 524} (2002) 265
[arXiv:hep-ph/0104071].

\bibitem{tables}
C.~W.~De Jager, H.~De Vries and C.~De Vries,
Atom.\ Data Nucl.\ Data Tabl.\ {\bf 14} (1974) 479.

\bibitem{Bialas}
A.~Bialas, M.~Bleszynski and W.~Czyz,
Nucl.\ Phys.\ B {\bf 111} (1976) 461.

\end{thebibliography}
\end{document}